\documentclass{iopart}

\usepackage{cite}
\usepackage{graphicx}

\usepackage{setstack}

\begin{document}

\title[Thermodynamic properties of plasma media]{Thermodynamic properties and electrical conductivity of strongly
correlated plasma media}

\author{
V S Filinov$^{1}$, P R Levashov$^{1}$, A V Bo\c{t}an$^{1}$, M
Bonitz$^{2}$ and V E Fortov$^{1}$}
\address{
$^1$Joint Institute for High Temperatures, Russian Academy of Sciences,\\
Izhorskaya 13 bldg 2, Moscow 125412, Russia\\
$^2$Christian-Albrechts-Universit{\"a}t zu Kiel, Institut f\"ur
Theoretische Physik und Astrophysik, Leibnizstrasse 15, 24098 Kiel,
Germany}
%\date{\today}

\begin{abstract}
%\boldmath
We study thermodynamic properties and the electrical
conductivity of dense hydrogen and deuterium using three methods:
classical reactive Monte Carlo (REMC), direct path integral Monte Carlo (PIMC)
and a quantum dynamics method in the Wigner representation of
quantum mechanics. We report the calculation of the deuterium
compression quasi-isentrope in good agreement with experiments. We
also solve the Wigner-Liouville equation of dense degenerate hydrogen calculating
the initial equilibrium state by the PIMC method.
The obtained particle trajectories determine the momentum-momentum
correlation functions and the electrical conductivity and are compared
with available theories and simulations.
\end{abstract}


%\maketitle

\section{Introduction}
% The very first letter is a 2 line initial drop letter followed
% by the rest of the first word in caps.
%
% form to use if the first word consists of a single letter:
% \IEEEPARstart{A}{demo} file is ....
%
% form to use if you need the single drop letter followed by
% normal text (unknown if ever used by IEEE):
% \IEEEPARstart{A}{}demo file is ....
%
% Some journals put the first two words in caps:
% \IEEEPARstart{T}{his demo} file is ....
%
% Here we have the typical use of a "T" for an initial drop letter
% and "HIS" in caps to complete the first word.
During the last decades significant efforts have been made to
investigate the thermophysical properties of dense plasmas. The
importance of this activity is mainly connected with the creation of
new experimental facilities. Powerful current generators and
ultrashort lasers are used for production of very high pressures
which cannot be reached in traditional explosive devices and
light-gas guns. Such experiments give valuable information about
various properties of strongly coupled plasmas. This allows one to
obtain a wide-range equation of state and to verify various
theoretical approaches and numerical methods. On the other hand
these results are of fundamental interest also for various
astrophysics and solid state physics applications.

Here we report new results on thermodynamic properties and
electrical conductivity of dense hydrogen and deuterium.
Thermodynamics is calculated by two approaches: the REMC method
\cite{Bezkrovniy:2004:PRE,Johnson:1994:MP} and the PIMC approach
\cite{Z-N-F:QMC1977}. Main attention is paid to the region of
hypothetical plasma phase transition (PPT). We also make a
comparison with recent experimental results on the quasi-isentropic
compression of deuterium.

To calculate the electrical conductivity we use quantum dynamics in the
Wigner representation of quantum mechanics. The Wigner-Liouville equation
is solved by a combination of molecular dynamics (MD) and MC methods.
The initial conditions are obtained using PIMC which yields thermodynamic
quantities such as the internal energy, pressure and pair distribution functions
in a wide range of density and temperature. To study the influence of the Coulomb interaction
on the dynamic properties of dense plasmas we apply the quantum dynamics in the
canonical ensemble at finite temperature and compute temporal momentum-momentum
correlation functions and their frequency-domain Fourier transforms. We discovered that these
quantities strongly depend on the plasma coupling parameter. For low
density and high temperature our numerical results agree well with
the Drude approximation and Silin's formula \cite{Silin:1964}, but with increasing
coupling parameter deviations grow.

\section{Simulation methods}

A complete description of the REMC method can be found in Refs.
\cite{Johnson:1994:MP, Bezkrovniy:2004:PRE}. Here we consider only
molecular hydrogen dissociation and recombination: $\mathrm{H}_{2}\Leftrightarrow 2\mathrm{H}$.
Ionization can be neglected at temperatures lower than the hydrogen
ground state energy ($\propto 13.6$~eV) and at moderate densities.
The effective pair potentials between different particle species are approximated
by Buckingham-EXP6 potentials, corrected at small distances \cite{Juranek:Redmer:2000:JCP}.
Our REMC simulations have been performed in the canonical ensemble for
hydrogen and deuterium. We use 3 types of MC moves: particle displacement,
molecular dissociation into atoms and recombination to a molecule. The expressions for
probabilities of the two last moves are given by the internal partition functions
of atoms and molecules $Z^{int}_A$, $Z^{int}_M$.
All electrons (in atoms and molecules) are assumed to be in the ground
state. Further, $Z^{int}_A$ contains only translational degrees of freedom,
$Z^{int}_M$ contains, in addition rotational and vibrational degrees of freedom.
For the latter we numerically solve the Schr\"{o}dinger equation
in the central-symmetric field, as described in Ref. \cite{Liu:2002},
which yields the energy levels $E_{nl}$ .

The PIMC method allows for first-principle simulations of dense plasmas at arbitrary coupling
and up to moderate degeneracy parameters, for details on our method, see \cite{Z-N-F:QMC1977,Filinov:2001:PPCF}. It has been
used for calculation of thermodynamic properties of hydrogen and
hydrogen-helium plasmas and electron-hole plasma in semiconductors.

Finally, we briefly describe our quantum dynamics (QMD) simulations of the conductivity.
Our starting point is the canonical ensemble-averaged time correlation
function \cite{zubar}
\begin{equation}
 C_{FA}(t) = \left\langle \hat{F}(0)\hat{A}%
(t)\right\rangle = Z^{-1}\mbox{Tr}\left\{\hat{F}
  e^{i\hat{H}t_c^{*}/\hbar} \hat{A} e^{-i\hat{H}t_c/\hbar}\right\},
\label{cfa}
\end{equation}
where $\hat{F}$ and $\hat{A}$ are operators of arbitrary observables and $\hat{H}$ is the Hamiltonian of the system which is the  sum of the kinetic $\hat{K}$ and the potential $\hat{U}$ energy
operators and the complex time is $t_c=t-i\hbar\beta /2$, and $\beta =1/k_BT$.
$Z=\mbox{Tr}\left\{e^{-\beta\hat{H}}\right\}$ is the partition
function. The Wigner representation of (\ref{cfa}) in a $\upsilon$--dimensional space is
\begin{equation*}
  C_{FA}(t) = (2\pi\hbar)^{-2\upsilon} \times{}
  \int\int d\mu_1d\mu_2 \,F(\mu_1)
\,A(\mu_2)\,W(\mu_1;\mu_2;t;i\hbar\beta),
\end{equation*}
where $\mu_i=(p_i,q_i), (i=1,2) $, and $p$ and $q$ comprise the
momenta and coordinates, respectively, of all particles. $A(\mu)$ and $F(\mu)$ denote Weyl's symbols
of the operators
$$
A(\mu)=\int d\xi e^{-i\frac{p\xi}{\hbar}} \left\langle
q-\frac{\xi}{2}\left |\hat{A}\right| q+\frac{\xi}{2}\right\rangle
$$
and $W(\mu_1;\mu_2;t;i\hbar\beta)$ is the spectral density expressed
as
\begin{eqnarray*}
&& W(\mu_1;\mu_2;t;i\hbar\beta) =
\frac{1}{Z}  \int\int d\xi_1d\xi_2
e^{i\frac{p_1\xi_1}\hbar}e^{i\frac{p_2\xi _2}\hbar}
\\
&\times&
\left\langle
q_1+\frac{\xi_1}{2}\left|e^{i\hat{H}t_c^{*}/\hbar}\right|
                     q_2-\frac{\xi_2}{2}\right\rangle  {}
\left\langle q_2+\frac{\xi_2}{2}\left|e^{-i\hat{H}t_c/\hbar}\right|
                     q_1-\frac{\xi_1}{2}\right\rangle.
\end{eqnarray*}

As has been proved in \cite{filmd1,filmd2},  $W$ obeys the following integral equation:
\begin{eqnarray}
W(&\mu_1;\mu_2;t;i\hbar\beta) =
\bar{W}(\bar{p}_0,\bar{q}_0;\tilde{p}_0,\tilde{q}_0;i\hbar \beta )
%  \bar{W}(\bar{\mu}_0;\tilde{\mu}_0;i\hbar\beta)
+ {}\nonumber\\
&\frac{1}{2}\int_0^td\tau\int ds
 W(\bar{p}_{\tau}-s,\bar{q}_{\tau};\tilde{p}_{\tau},\tilde{q}_{\tau}
   \tau ;i\hbar\beta) \varpi (s,\bar{q}_{\tau}) - {}\\
&\frac{1}{2}\int_0^td\tau\int ds
 W(\bar{p}_{\tau},\bar{q}_{\tau};\tilde{p}_{\tau}-s,\tilde{q}_{\tau};
 \tau ;i\hbar\beta) \,\varpi (s,\tilde{q}_{\tau})\nonumber,
 \label{eq:WL}
\end{eqnarray}
where $\bar{W}(\mu_1;\mu_2;i\hbar\beta )\equiv
W(\mu_1;\mu_2;0;i\hbar\beta)$ is the initial condition equation,
which can be presented in the form of a finite difference
approximation of the Feynman path
integral~\cite{filmd1,filmd2}.
The expression for $W$  has to be generalized to account for the spin
effects. This gives rise to an additional spin part of the initial density
matrix, e.g. \cite{Filinov:2001:PPCF}. Also, to
improve the simulation accuracy the pair interactions
$U_{ab}$, are replaced by an effective quantum potential $U_{ab}^{\rm eff}$,
such as the Kelbg potential~\cite{Ke63}. For details we
refer to Refs.\cite{Z-N-F:QMC1977,prtthr,filinov76,znf76,Filinov:2001:PPCF}, for
recent applications of the PIMC approach to correlated Coulomb systems,
cf. \cite{Filinov:2000,Filinov:2000:PLA,Levashov:2001,Filinovjr:2001:PRL}.

The solution of the integral equation (\ref{eq:WL}) can be
represented by an iteration series
$$
W^t = \bar{W}^t+K_\tau ^tW^\tau = \bar{W}^t+K_{\tau
_1}^t\bar{W}^{\tau_1} +\dots,
$$
where $\bar{W}^t$ and $\bar{W}^{\tau _1}$ are the initial quantum
spectral densities evolving classically during time intervals
$[0,t]$ and $[0,\tau_1]$, respectively, whereas
$K_{\tau_i}^{\tau_{i+1}}$ are operators that govern the propagation
from time $\tau_i$ to $\tau_{i+1}$, see e.g. \cite{filinov_prb}.
Thus the time correlation function becomes
\begin{eqnarray*}
  C_{FA}(t) = \left(\phi|W^t\right) =
 \left(\phi|\bar{W}^t\right)
+\left(\phi|K_{\tau_1}^t\bar{W}^{\tau_1}\right) +\dots
\end{eqnarray*}
where $\phi(\mu_1;\mu_2)\equiv F(\mu_1)A(\mu_2)$ and the parentheses
$\left(\dots|\dots\right)$ denote integration over the phase space
$\{\mu_1;\mu_2\}$.

The iteration series for $C_{FA}(t)$ can be efficiently computed
using MC methods. We have developed a MC scheme which provides
domain sampling of the terms giving the main contribution to the
iteration series, cf. \cite{filmd1,filmd2}. For simplicity, in this
work, we take into account only the first term of iteration series,
which is related to the propagation of the initial quantum
distribution $\bar{W}$ according to the Hamiltonian equations of
motion. This term, however, does not describe pure classical
dynamics but accounts for quantum effects \cite{filmd3} and, in
fact, contains arbitrarily high powers of the Planck's constant.

\section{Numerical results}

\subsection{Deuterium compression isentrope}
To calculate an isentrope one has to determine the entropy
which is defined by: $S=-[\partial F(T,V,N)/\partial T]_V$,
where $F=\sum_{i=1}^2N_{i}\mu_{i} - PV$ is
the free energy of a two-component system of atoms and molecules. The chemical
potentials $\mu_i$ of each component are obtained with the the test particle
method \cite{Widom:1963}: $\mu=\mu^{id}+\mu^{r}$,
where $\mu^{id}$ is the ideal gas chemical potential:
$\mu^{id}=k_BT\log(\Lambda^3N/V)$ and $\Lambda$ is the de Broglie
wavelength of a particle. The residual chemical potential $\mu^{r}$ can be evaluated as
\cite{Arrieta:1991}:
$$
  \mu^{r}=-k_BT\log[L_a\exp(-\Delta U/k_BT)].
$$
Here $\Delta U$ denotes the change of configurational energy produced by
the insertion of one additional particle and $L_a$ is the ratio of
allowed (nonoverlapped) insertion intervals along trajectories which
traverse (parallel to any of the axes) the simulation box from side
to side, to the length of the box \cite{Arrieta:1991}.
%In the Widom's method the particle is inserted into some random point of the computational cell; in this work we use random insertion trajectories \cite{Arrieta:1991} which traverse (parallel to any of the axes) the simulation volume from side to side. Thus we determine the allowed (nonoverlapped) insertion intervals along the trajectories and calculate the total allowed length $L_a$ for a given configuration of particles.
The test particle is then inserted randomly into some point in the
allowed intervals and the change in configurational energy $\Delta
U$ is evaluated. The main advantage of this approach is the
possibility of calculation of chemical potentials at high densities,
where the usual test particle method tends to fail. The chemical
potentials are calculated separately for atoms and molecules.

The isentrope can also be calculated by using Zel'dovich's method
\cite{Zel'dovich:1957, Fortov:1970}. From the first law of
thermodynamics the characteristic equation for the temperature along
the isentrope can be derived:
$$
\frac{dT}{dV}=-\frac{T}{(\partial E/\partial P)_{V}}.
$$
We integrate this equation with the initial condition corresponding
to an experimental point at low pressure taking the temperature at
this point to be close to that from the Widom's test
particle method. The coefficient $(\partial E /
\partial P)_V$ is obtained from the interpolation functions $E(T,V)$
and $P(T,V)$, which are constructed from the REMC calculation
 on the grid of isotherms and isochores covering the
experimental isentrope.

Calculations were performed in a cubic simulation box with periodic
boundary conditions, and with a cutoff radius equal to half of the
box length. The initial particle configuration was an fcc lattice for every input density
$\rho=N_{\mathrm{H}}m_{\mathrm{H}}/V+N_{\mathrm{H}_{2}}m_{\mathrm{H}_{2}}/V$
with $N_{\mathrm{H}}=N_{\mathrm{H}_{2}}=250$. The system was
equilibrated for $2\cdot 10^7$ steps, and additional $10^7$ steps were
used for the calculation of thermodynamic values. Averaging of 20
blocks was used to calculate the statistical error, which did not
exceeded $2\%$ for pressure and energy.

\begin{figure}[h]
\centering
\includegraphics[width=0.42\columnwidth]{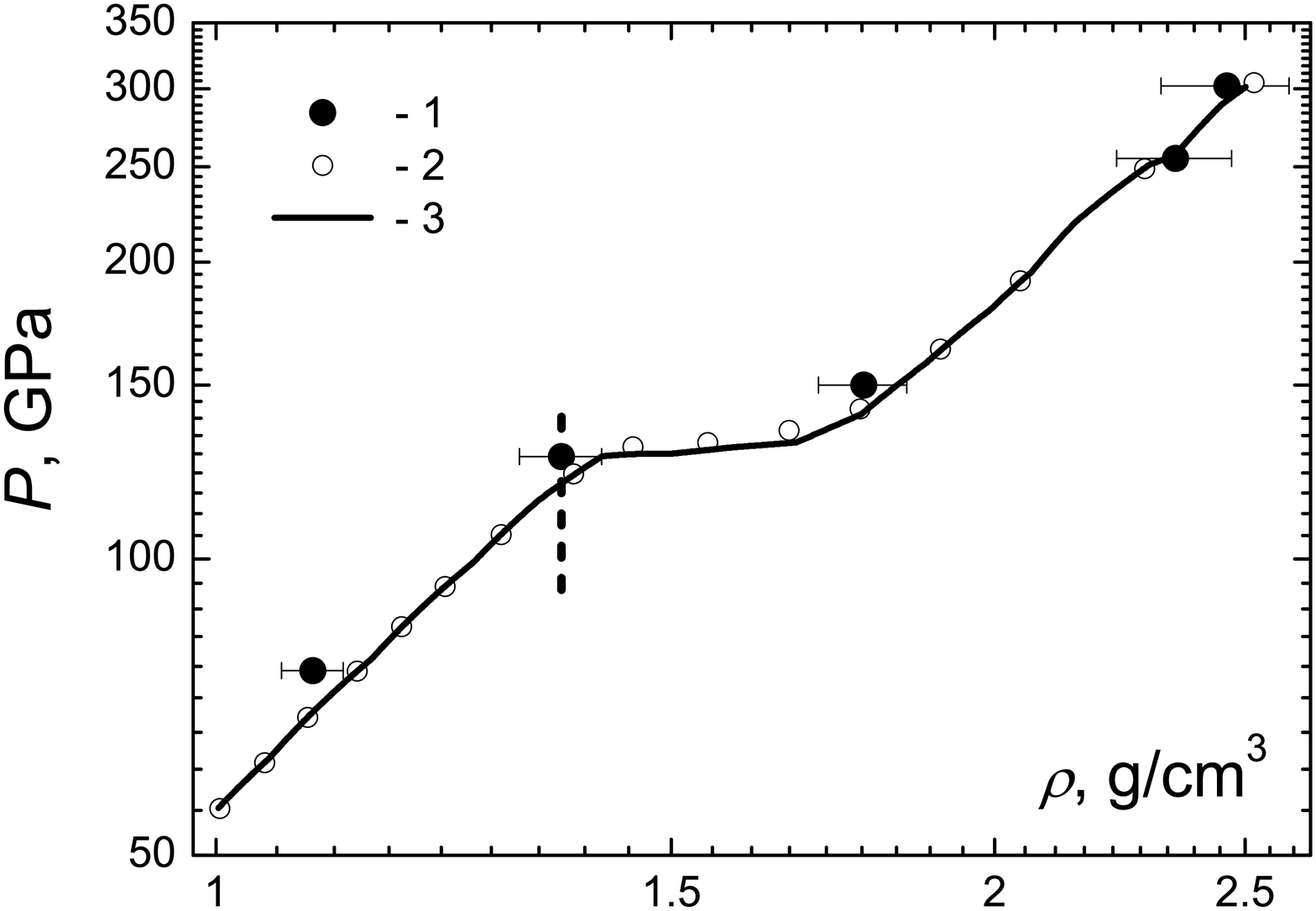}% Here is how to import EPS art
\hspace{0.08\columnwidth}
\includegraphics[width=0.42\columnwidth]{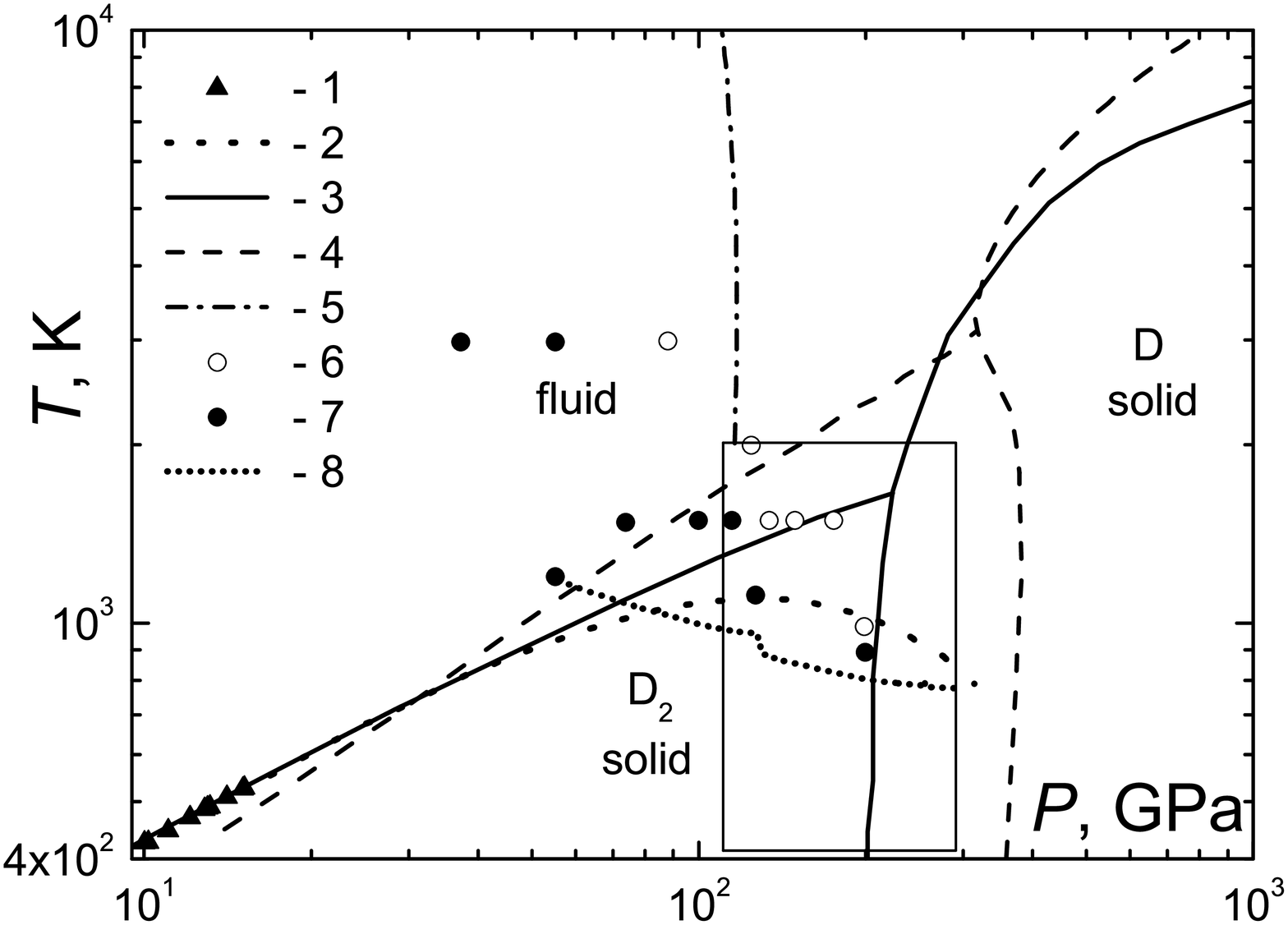}% Here is how to import EPS art
\caption{\label{isentrope} Deuterium quasi-isentropic compression.
Experiment: 1~---\cite{Fortov:2007:PRL}.
 This work: 2~--- Widom's test particle method, 3~--- Zel'dovich's method. Vertical
 line shows the density at which a sharp electircal conductivity rise
 is observed experimentally \cite{Weir:1996,Ternovoi:1999:PB}.}
\caption{\label{phases} Deuterium phase diagram. Melting curve:
1~--- experiment \cite{Datchi:2000}; 2~--- extrapolation of
experiment \cite{Datchi:2000}. Theoretical phase diagram:
3~---\cite{Kerley:1983}; 4~---\cite{Kopyshev}. Boundary of the
possible phase transition: 5~---\cite{Beule:1999}. Atomic fluid:
6~---\cite{Bonev:2004,Scandolo:2003}. Molecular liquid:
7~---\cite{Bonev:2004,Scandolo:2003}. This work: 8~--- isentrope;
the black rectangle shows the region of slow convergence to the
equilibrium state.}
\end{figure}

The results for three shock Hugoniots of gaseous
deuterium with three different initial densities ($\rho_0 =
0.1335$~g/cm$^3$, $P_0 = 1.57$~GPa; $\rho_0 = 0.153$~g/cm$^3$, $P_0
= 2.03$~GPa; $\rho_0 = 0.171$~g/cm$^3$, $P_0 = 2.72$~GPa) obtained by REMC
and DPIMC methods can be found elsewhere \cite{Botan:2008}. Here we
present our recent results concerning the calculation of deuterium
quasi-isentrope of compression. A dramatic increase of the
conductivity of deuterium by 4--5 orders of magnitude \cite{Weir:1996}
at pressures $\sim$1~Mbar and densities about 1~g/cm$^3$ indicates
that one might also expect peculiarities in the thermodynamic
properties. Indeed, there are experimental results on
quasi-isentropic compression of deuterium in a cylindrical explosive
chamber which show a 30\% density jump at a pressure of about 1.4 Mbar
\cite{Fortov:2007:PRL}. Using the deuterium free energy from
 REMC and Widom's test particle methods we calculated
the compression isentrope of deuterium. The results are shown in
Fig.~\ref{isentrope} and compared to experimental data
\cite{Fortov:2007:PRL}. The excellent agreement proves that no special
assumptions about a phase transition are needed to explain such a
remarkable behavior of the experimental points, apparantly dissociation effects
mostly contribute to this phenomenon.

A comparison of our results with other theoretical and
computational methods is shown in the phase diagram Fig.\ref{phases}.
The temperature and density jump along the isentrope is close to
the predicted boundary of the deuterium phase transition from the
molecular to the atomic phase \cite{Beule:1999,Bonev:2004,
Scandolo:2003}. The rectangle shows the region of slow convergence of the simulations
to the equilibrium state which might be an indication of a
thermodynamic instability. Our preliminary analysis shows that this
state can be stable (up to $\approx 2\cdot 10^9$ Monte-Carlo steps
are required), but to investigate the possibility of phase
separation at these conditions one needs to apply special Monte
Carlo methods.

\subsection{Quantum momentum-momentum correlation functions}

We now compute the dynamic conductivity of a strongly coupled
hydrogen plasma. The results obtained were practically insensitive
to the variation of the whole number of particles in the Monte Carlo
cell from 30 up to 60 and also to the number $n$ of high temperature
density matrices in the path integral representation of the initial state
which ranged from $n=20$ to 40. Estimates of the average statistical error gave a
value of the order  5--7\%.
According to the Kubo formula \cite{zubar} our calculations include
two stages: (i)  generation of the initial conditions
(configuration of protons and electrons) in the canonical ensemble
with the probability being proportional to the quantum density matrix and
(ii), generation of the dynamic trajectories in phase space,
starting from these initial configurations.

\begin{figure}[htb]
\centering
\includegraphics[width=0.32\columnwidth]{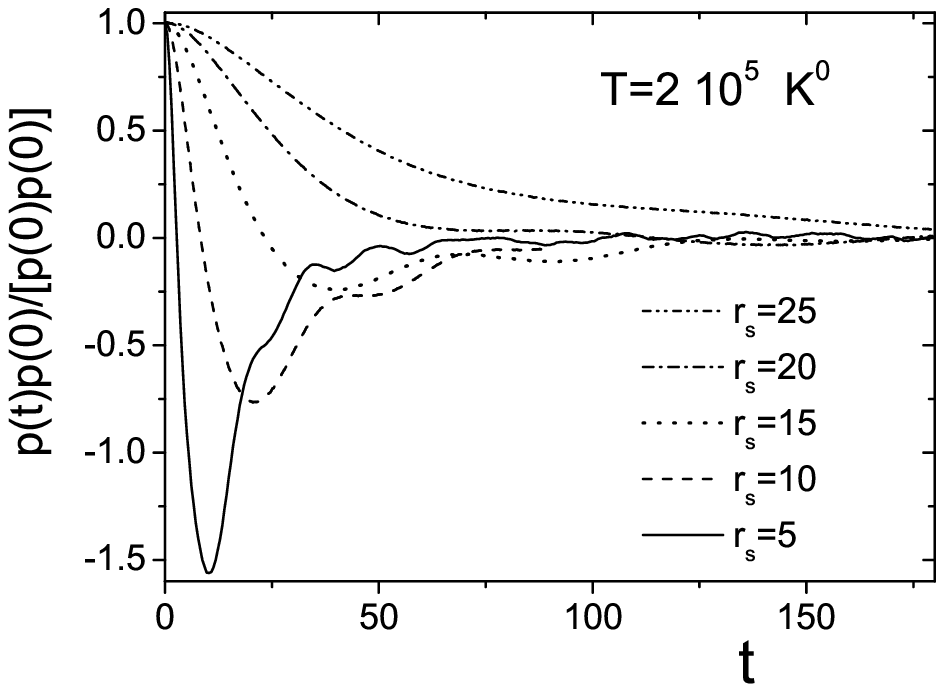}
%\hspace*{0.5cm} \vspace*{-0.2cm}
\includegraphics[width=0.32\columnwidth]{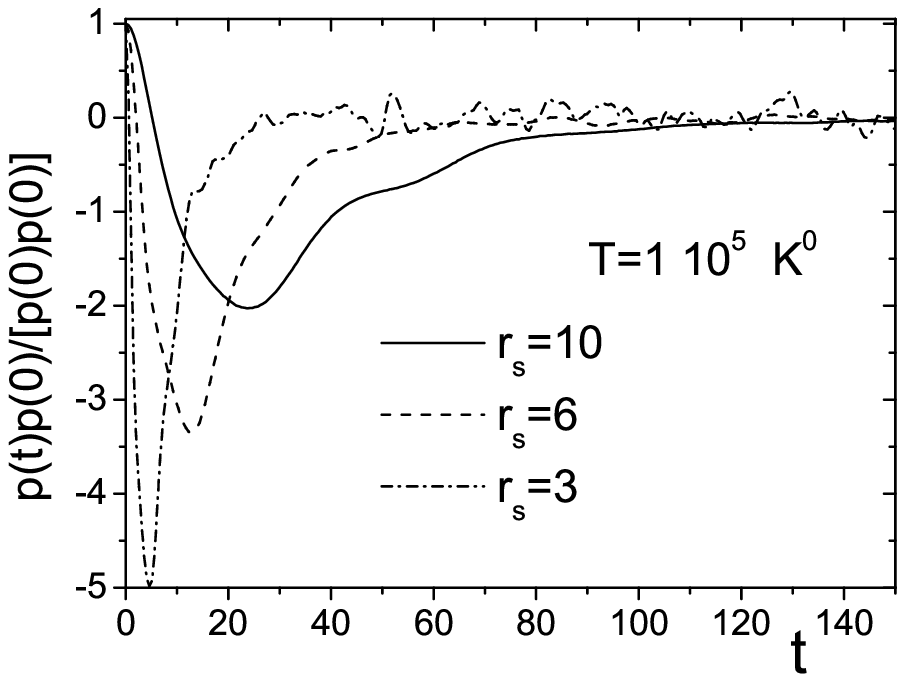}
\includegraphics[width=0.32\columnwidth]{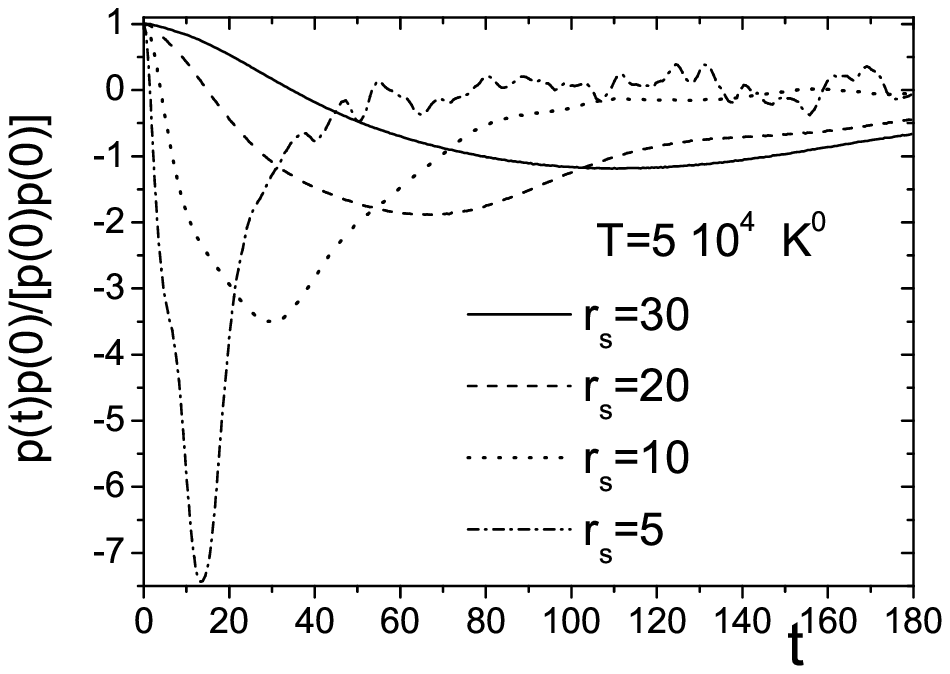}
\caption[]{Typical MMCF in canonical ensemble for different
densities ($r_s$) and three temperatures: $T=200\,000$~K (left),
$T=100\,000$~K (central) and $T=50\,000$~K (right). Time is
presented in atomic units.} \label{ppcor}
\end{figure}

First we discuss the momentum-momentum autocorrelation functions
(MMCF) which are shown in Fig.~\ref{ppcor} for various temperatures
and densities.
%varying over one order of magnitude.
The plasma density is characterized by the Brueckner parameter $r_s$
defined as the ratio of the mean interparticle distance
$d=(\frac{3}{4\pi (n_e+n_p)})^{1/3}$ to the Bohr radius $a_B$, where
$n_e$ and $n_p$ are the electron and proton densities. The monotonic
decay of the MMCF at low density transforms into aperiodic
oscillations at high densities. The tails of the MMCF clearly show
collective plasma oscillations. The damping time of the initial
decay of the MMCF is strongly affected by variations of density and
temperature. At constant density the decay time is at least two
times smaller for $T = 200\,000$~K compared to  $T = 50\,000$~K. As
it follows from Fig.~\ref{ppcor} the damping time is also sensitive
to the density. The damping time of the MMCF increases when the
density increases. The physical reason is the tendency towards
ordering of the charges.

\subsection{Electrical conductivity}

\begin{figure}[htb]
\centering
\includegraphics[width=0.32\columnwidth]{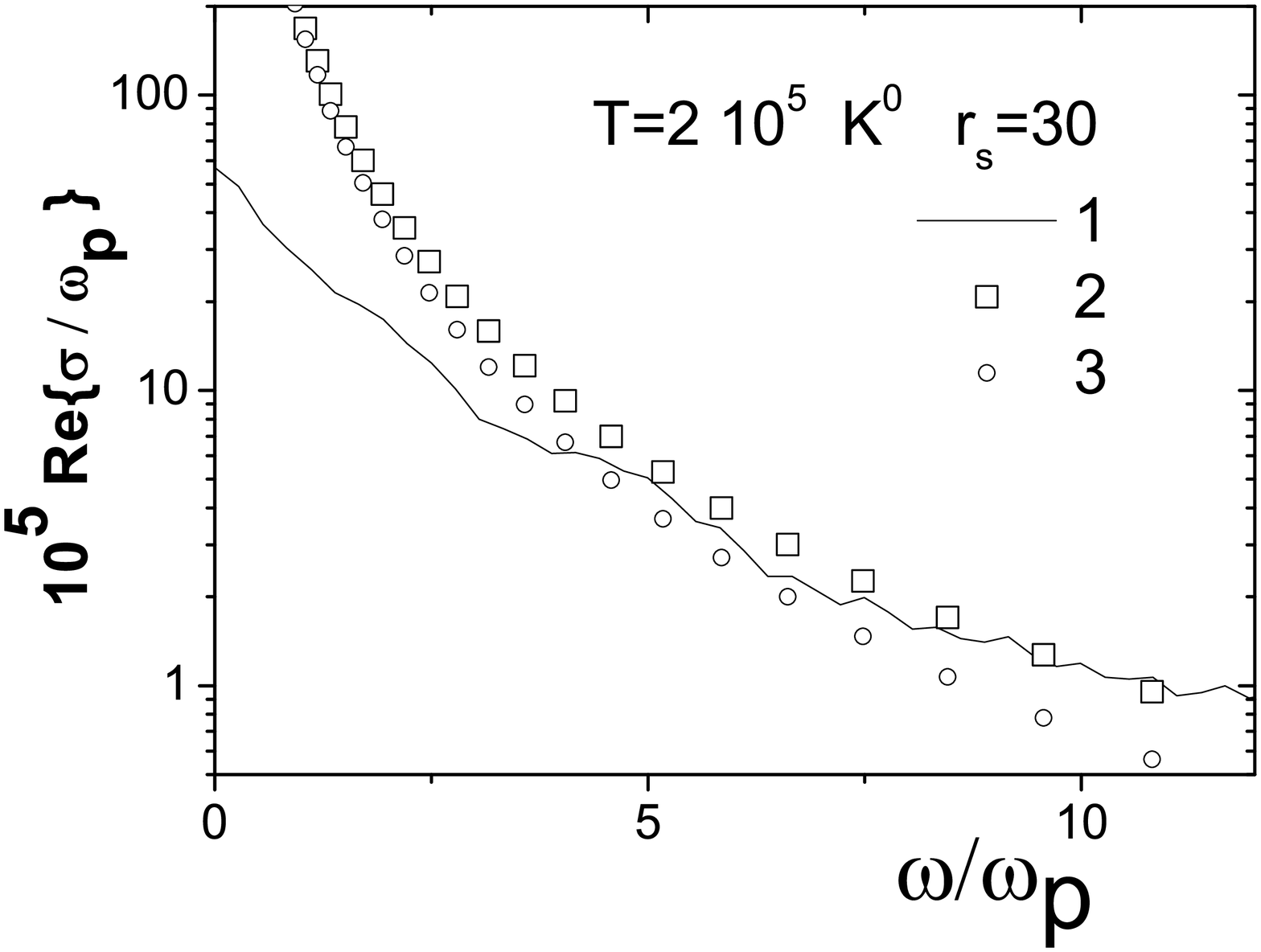}
%\hspace*{0.5cm} \vspace*{-0.2cm}
\includegraphics[width=0.32\columnwidth]{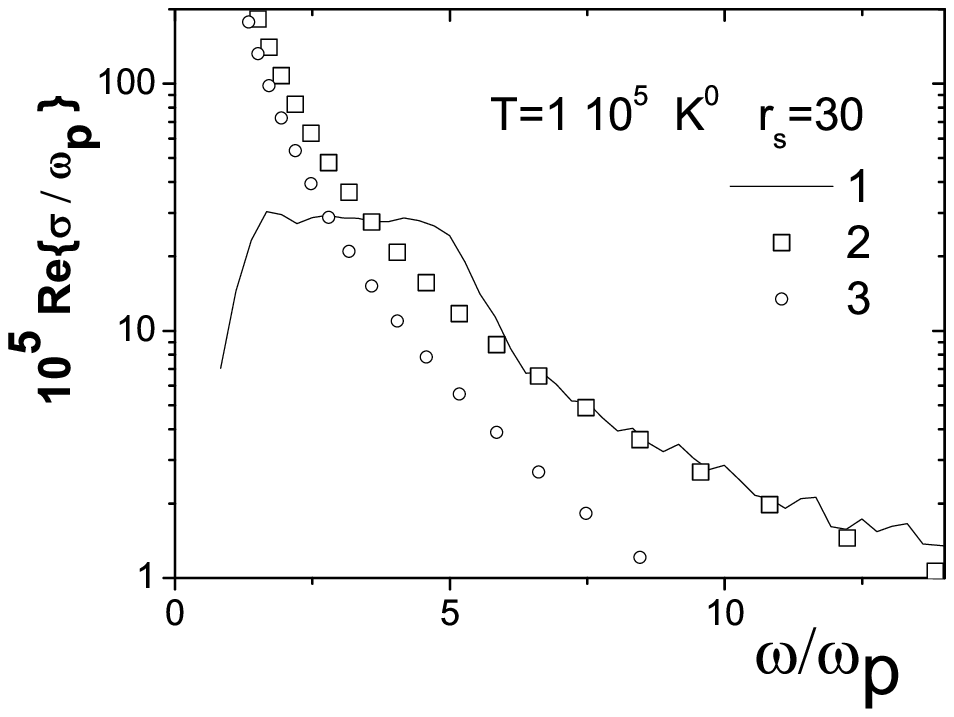}
\includegraphics[width=0.32\columnwidth]{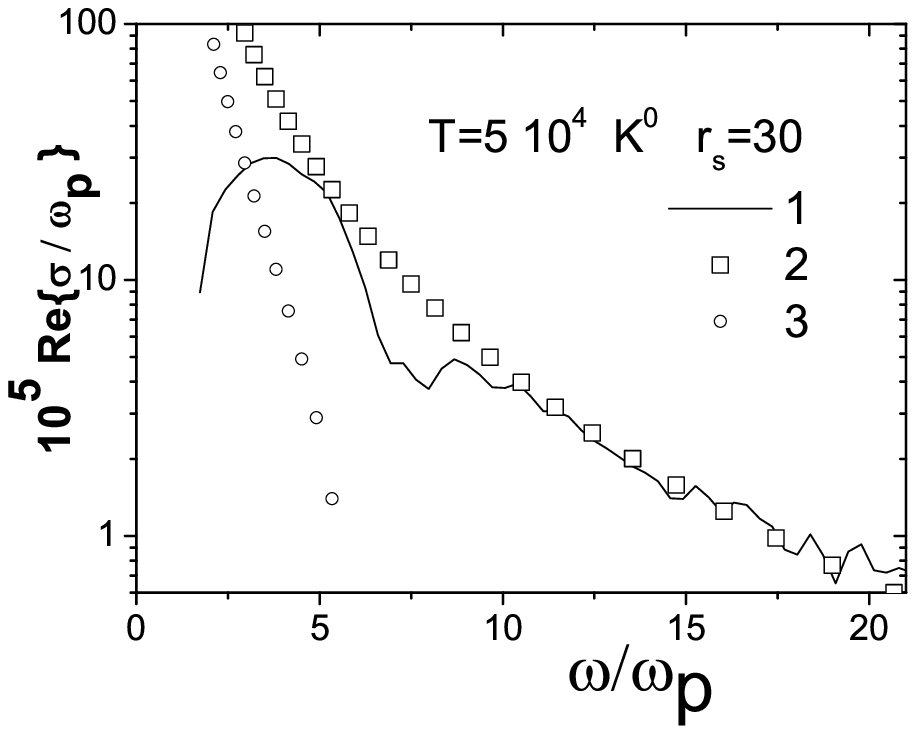}
\includegraphics[width=0.32\columnwidth]{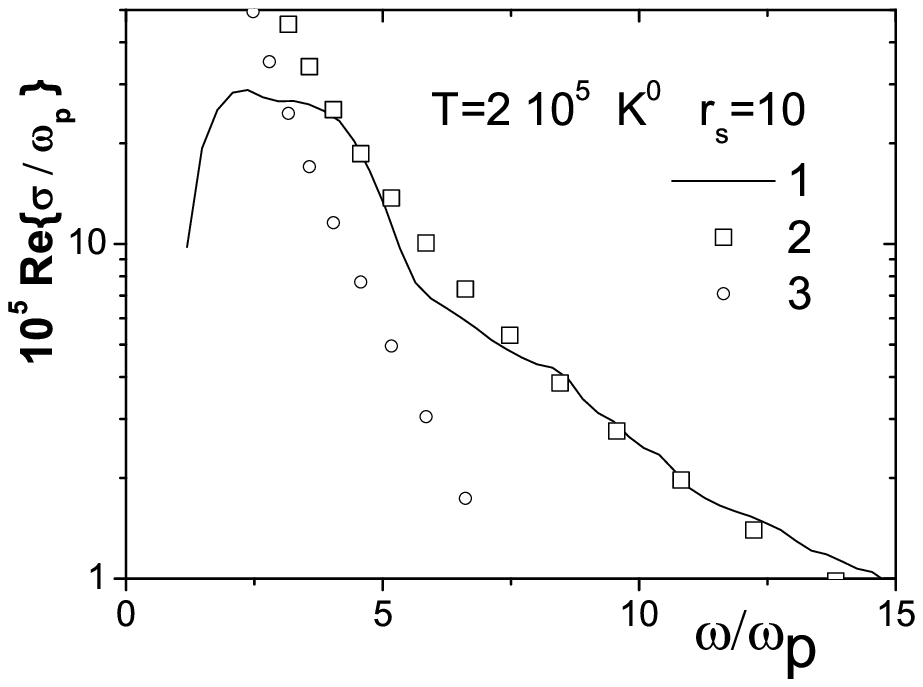}
\includegraphics[width=0.32\columnwidth]{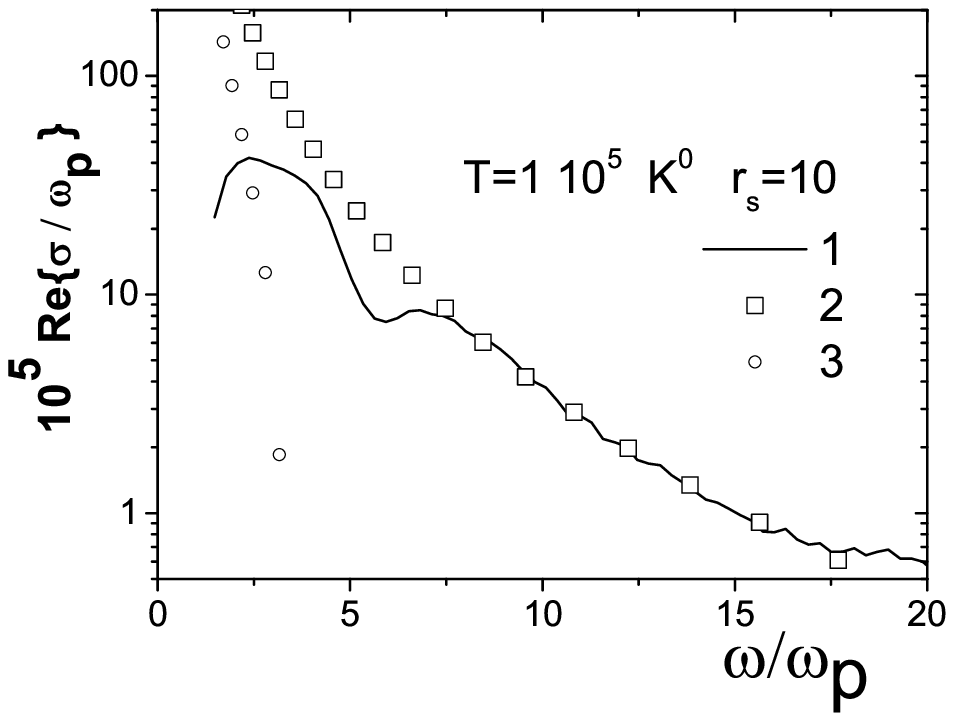}
%\hspace*{0.5cm} \vspace*{-0.2cm}
%\hspace*{0.5cm} \vspace*{-0.2cm}
%\hspace*{0.5cm} \vspace*{-0.2cm}
\includegraphics[width=0.32\columnwidth]{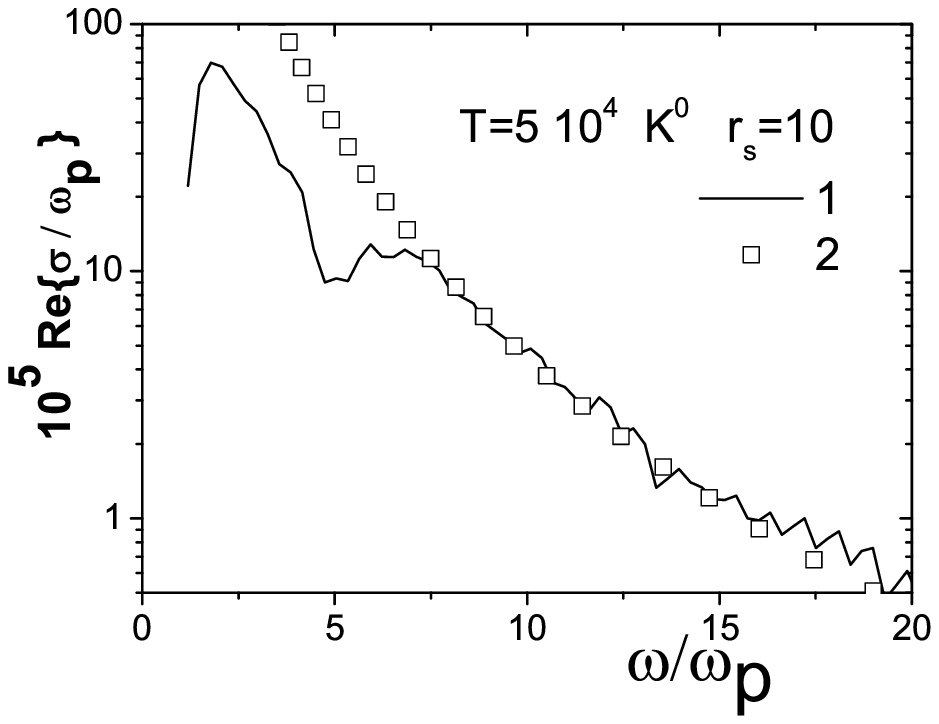}
\caption{Real part of the Fourier transform of the MMCF (line 1) for
densities related to three temperatures: $T=200\,000$~K (left
column), $T=100\,000$~K (central column) and $T=50\,000$~K (right
column) and to $r_s=30$ (top row) and $r_s=10$ (bottom row). Points
$2$ and $3$ present analytical results obtained according to
\cite{Bornath:2000,Silin:1964} respectively. Frequency and scaled
dynamic conductivity is given in units of plasma frequency.}
\label{Sig30} \centering
\end{figure}

Figure~\ref{Sig30} presents the real  part of the diagonal elements
of the electrical conductivity tensor versus frequency computed from the
real part of the Fourier transform of the MMCF which characterizes the Ohmic absorption of
electromagnetic energy.
Collective plasma oscillations give the main contribution in the
region of $\omega / \omega_p\sim 1$, where $\omega_p^2 = 4\pi n_e e^2 / m_e$
is the plasma frequency.  Reliable
numerical data in this region require very long dynamic
trajectories. Their initial parts are presented in
Fig.~\ref{ppcor}. The high frequency tails of the dynamic conductivities
coincide with analytical Drude-like expressions for fully ionized
hydrogen plasma obtained in \cite{Silin:1964,Mihajlov:2001}. For low
frequency analytical estimations are going to infinity and this is
the reason of discrepancy between numerical results and analytical
estimations. With increasing plasma density non-monotonic behavior of
the dynamic conductivity in the region of
several plasma frequencies is observed. These oscillations can be an indication
of the transparency window (low absorption coefficient) of the strongly coupled hydrogen plasma.
%discussed in \cite{filmd?,filmd??}.

%\begin{figure}[htb]
%\begin{center}
%\includegraphics[width=4.2cm,clip=true]{lgrer10t200.eps}
%%\hspace*{0.5cm} \vspace*{-0.2cm}
%\includegraphics[width=4.2cm,clip=true]{lgrer10t100.eps}
%%\hspace*{0.5cm} \vspace*{-0.2cm}
%\includegraphics[width=4.2cm,clip=true]{lgrer10t50.eps}
%\caption{Real part of the Fourier transform of the MMCF for
%densities related to $r_s=10$ and three temperatures: $T=200\ 000
%K^o$ (left panel),$T=100\ 000 K^o$ (central panel) and $T=50\ 000
%K^o$ (right panel).  Notatiopns are explained on
%Fig.~\ref{Sig30}.} \label{Sig10}
%\end{center}
%\end{figure}

\begin{figure}[htb]
\begin{center}
\includegraphics[width=0.32\columnwidth]{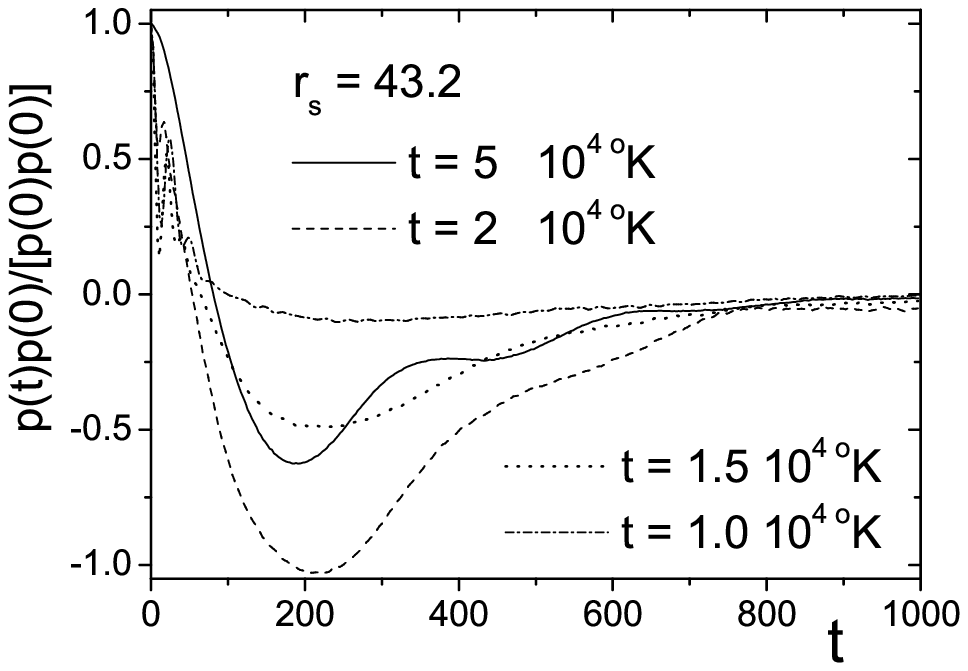}
%\hspace*{0.5cm} \vspace*{-0.2cm}
\includegraphics[width=0.32\columnwidth]{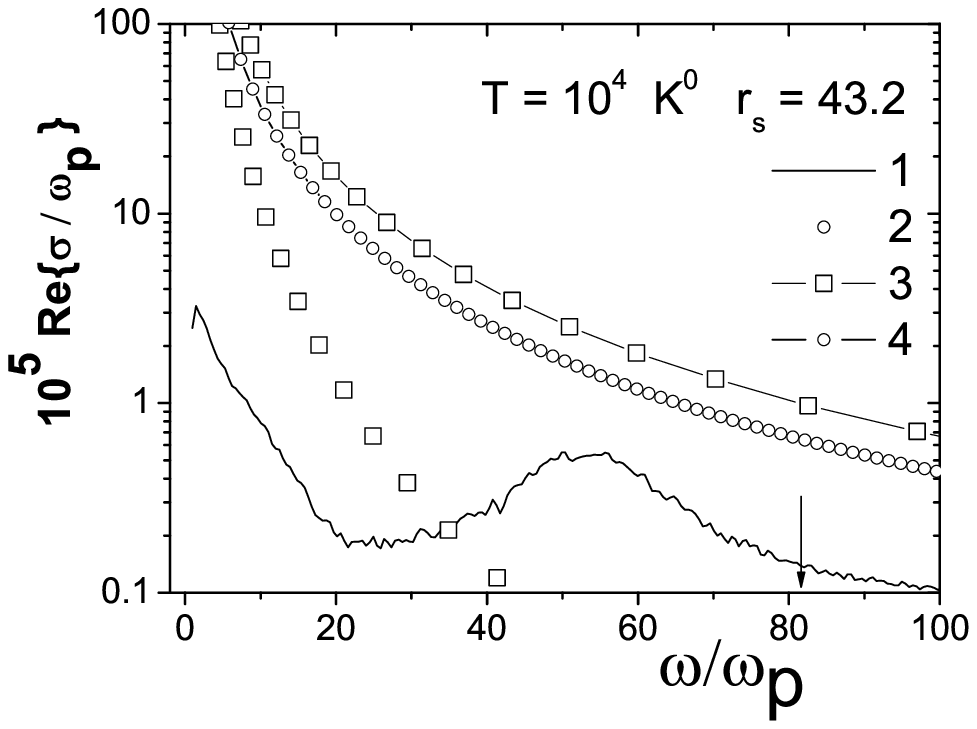}
%\hspace*{0.5cm} \vspace*{-0.2cm}
\includegraphics[width=0.32\columnwidth]{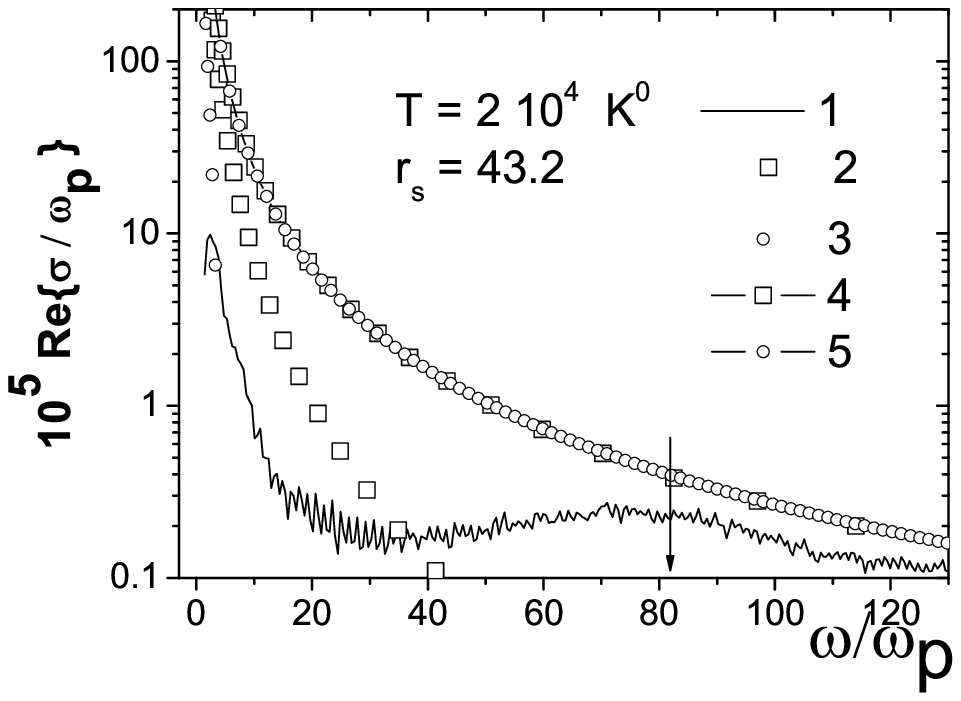}
\caption{Typical MMCF in canonical ensemble for $r_s=43.2$ and four
temperatures: $T=50\,000$~K, $T=20\,000$~K, $T=15\,000$~K and
$T=10\,000$~K, time is presented in atomic units (left figure). Real
part of the Fourier transform of MMCF (line 1) for density related
to $r_s=43.2$ and temperatures $T=10\,000$~K (central figure) and
$T=20\,000$~K (right figure). Data points 2--5 present analytical
results obtained according to
\cite{Bornath:2000,Silin:1964,Mihajlov:2001,AdmTkach1} respectively.
Arrow relates to the condition $Ry=\hbar \omega$. } \label{Rs43}
\end{center}
\end{figure}

The agreement with Drude-like formulas for weakly coupled plasma is
due to the fact that the main contribution to the high frequency
region comes from the fast trajectories with high (virtual) energy.
This means that interaction of electrons with each other and protons
only weakly disturbs the behavior of the high-frequency tails of
dynamic conductivity in comparison with ideal plasma. Now let us
consider the dynamic conductivity at very low plasma density, namely
when the Brueckner parameter $r_s$ is equal to $43.2$.
Fig.~\ref{Rs43} presents MMCF and dynamic conductivity in a wide
region of temperatures from $T=50\,000$~K up to $T=10\,000$~K. At
temperatures lower than $T=50\,000$~K the plasma consists mainly of
atoms. As a consequence the initial fast decay of the MMCF is
modulated by the high frequency oscillations related to the motion
of electrons inside the atoms. So the high frequency tail of the
dynamic conductivity has a new maximum associated with the condition
$Ry=\hbar \omega$. As it follows from Fig.~\ref{Rs43} the position
of this peak strongly depends on temperature and is shifted to lower
frequency when temperature decreases. The physical reason is the
growth of the energy levels population of hydrogen atoms with the
increase of temperature. Analytical estimations  for fully ionized
plasma which are also presented  in Fig.~\ref{Rs43} give essentially
larger values of dynamic conductivity.

\section{Conclusion}
We have used three approaches for the investigation of thermodynamic
properties and electrical conductivity of dense hydrogen and
deuterium plasma. Using two different methods of isentrope
reconstruction from simulation results we obtained very good
agreement with experimental data on quasi-isentropic compression of
deuterium. We also applied the quantum dynamics approach to hydrogen
plasma in a wide region of density and temperature. Calculating the
MMCF we determined the dynamic electrical conductivity and compared
the results with available theories. Our results show a strong
dependence on the plasma coupling parameter. For low density and
high temperature the numerical results agree well with the Drude
approximation, while at higher values of the coupling parameter we
observe a strong deviation of the frequency dependent conductivity
and permittivity from low density and high temperature
approximations.

% use section* for acknowledgement
\section*{Acknowledgment}
The authors thank H.~Fehske for stimulating discussions and valuable
notes. V. Filinov acknowledges the hospitality of the of the
Institut f\"ur Theoretische Physik und Astrophysik of the
Christian-Albrechts-Universit{\"a}t zu Kiel.

\section*{References}

%\bibliographystyle{iopart-num}
%\bibliography{refs}

\hyphenation{Post-Script Sprin-ger} \providecommand{\newblock}{}

\end{document}